\documentclass[sigconf, authorversion, nonacm]{acmart}

\AtBeginDocument{%
  }

\usepackage{amsmath}
\usepackage{graphics}
\usepackage{color}
\usepackage{fontawesome}
\usepackage{xspace}
\usepackage{soul}
\usepackage{siunitx}
\usepackage[flushleft]{threeparttable}
\usepackage[title]{appendix}
\usepackage{enumitem}
\usepackage{balance}
\usepackage{xspace}
\usepackage{xcolor}
\usepackage{color, colortbl, soul}
\usepackage{tikz}
\usepackage{pifont}

\def\aka{\textit{a.k.a.}\xspace}

\setcopyright{acmlicensed}
\copyrightyear{2018}
\acmYear{2018}
\acmDOI{XXXXXXX.XXXXXXX}

\acmConference[Conference acronym 'XX]{Make sure to enter the correct
  conference title from your rights confirmation emai}{June 03--05,
  2018}{Woodstock, NY}
\acmISBN{978-1-4503-XXXX-X/18/06}



\begin{document}

\title{VR for Acupuncture? Exploring Needs and Opportunities for Acupuncture Training and Treatment in Virtual Reality}

\author{Menghe Zhang}
\email{mez071@ucsd.edu}
\affiliation{%
  \department{Computer Science and Engineering}
  \institution{University of California San Diego}
  \city{La Jolla}
  \state{CA}
  \country{United States}
}

\author{Chen Chen}
\email{chenchen@ucsd.edu}
\orcid{0000-0001-7179-0861}
\affiliation{%
  \department{Computer Science and Engineering}
  \institution{University of California San Diego}
  \city{La Jolla}
  \state{CA}
  \country{United States}
}

\author{Matin Yarmand}
\email{myarmand@ucsd.edu}
\orcid{0000-0002-2353-3462}
\affiliation{%
  \department{Computer Science and Engineering}
  \institution{University of California San Diego}
  \city{La Jolla}
  \state{CA}
  \country{United States}
}

\author{Nadir Weibel}
\email{weibel@ucsd.edu}
\orcid{0000-0002-3457-4227}
\affiliation{%
  \department{Computer Science and Engineering}
  \institution{University of California San Diego}
  \city{La Jolla}
  \state{CA}
  \country{United States}
}

\begin{abstract}
Acupuncture is a form of medicine that involves inserting needles into targeted areas of the body and requires knowledge of both \textbf{T}raditional \textbf{C}hinese \textbf{M}edicine (TCM) and \textbf{E}vidence-\textbf{B}ased \textbf{M}edicine (EBM).
The process of acquiring such knowledge and using it for practical treatment is challenging due to the need for a deep understanding of human anatomy and the ability to apply both TCM and EBM approaches.
Visual aids have been introduced to aid in understanding the alignment of acupuncture points with key elements of the human body, and are indispensable tools for both learners and expert acupuncturists. However, they are often not enough to enable effective practice and fail to fully support the learning process.
Novel approaches based on immersive visualization and \textbf{V}irtual \textbf{R}eality (VR) have shown promise in many healthcare settings due to their unique advantages in terms of realism and interactions, but it is still unknown whether and how VR can possibly be beneficial to acupuncture training and treatment. 
Following participatory design protocols such as observations and semi-structured interviews with eight doctors and nine students, we explore the needs and pain points of current acupuncture workflows at the intersection of EBM and TCM in China and the United States. We highlight opportunities for introducing VR in today's acupuncture training and treatment workflows, and discuss two design approaches that build on $11$~specific challenges spanning education, diagnosis, treatment, and communication.
\end{abstract}

\settopmatter{printfolios=true}
\maketitle

\section{Introduction}
Acupuncture~\citep{kaptchuk2002acupuncture} is an important form of alternative and integrative medicine that involves precisely inserting thin needles into targeted body areas. It has long been used in clinical practices around the world. The key component of acupuncture treatment is identifying and stimulating anatomic sites (\aka~acupuncture points, or acupoints).

Compared to general medical practices, acupuncture treatment and training face unique challenges. Unlike Western medicine, acupuncture relies on the practitioner's experience in comprehending the intricate interplay between human anatomy and the overall health of the patient in a holistic manner, drawing from the principles of \textbf{T}raditional \textbf{C}hinese \textbf{M}edicine (TCM)~\citep{white2009western}. In addition, to ensure effective and safe treatments, it also requires expertise in \textbf{E}vidence-\textbf{B}ased \textbf{M}edicine~(EBM), which integrates the best available scientific evidence with clinical expertise and patient values. However, combining the principles of TCM with EBM approaches can be challenging for novices without practical experience in understanding these intricate connections. Moreover, acupoints are not precisely described in anatomical terms due to limited research on the underlying mechanisms. This presents a challenge in ensuring consistent quality of care. Furthermore, the current acupuncture training process heavily relies on texts and atlases that do not fully capture the three-dimensional nature and nuances of real-world human anatomy.

\textbf{V}irtual \textbf{R}eality (VR) has emerged as a promising solution for diverse medical contexts~\citep{liu2022virtual} and medical training programs~\citep{nicholson2006can,pottle2019virtual,makinen2022user} due to the unique advantages of natural perceptions of 3D information and the support for more intuitive interactions. While a few existing products, such as AcuMap~\citep{maiacumap}, have made attempts to integrate acupuncture training workflows into virtual reality, it is still unclear {\it why} and {\it how} the VR can effectively facilitate acupuncture treatment and training workflow.

In this work, we investigate the potential benefits of using VR to enhance acupuncture treatment and training workflows from the perspective of both practitioners and students. Following semi-structured interviews with eight acupuncture doctors and nine students from professional medical institutes located in China and the United States, we identified $11$ challenges spanning education, treatment, and cooperation. We present our initial insights and discuss design opportunities to address the challenges identified by participants across three levels of expertise, encompassing both training and treatment. Through this process, we identified three guiding themes that apply to both aspects: the acquisition of fundamental knowledge, the accumulation of clinical experience, and the facilitation of demonstration and collaboration.

In summary, this paper highlights two main contributions to the field of Human-Computer Interaction (HCI) and the design of Virtual Reality systems for acupuncture training and treatment:

\begin{itemize}[leftmargin=*]
    \item {\bf Need--Finding.} Through a need--finding process, we examine challenges faced by acupuncture students and practitioners in their daily routines.
    \item {\bf Design Considerations.} We propose two potential design principles for VR systems to address the challenges emerged in our need--findings.    
\end{itemize}

We believe that the findings of this work will offer critical insights to inform future researchers and practitioners who aim to design and develop collaborative VR applications in support of acupuncture training and treatment workflows.
\section{Background and Related Work}
\label{sec:background}

In this section, we first introduce the existing processes of acupuncture training and treatment. Then, we give an overview of the literature on immersive visualizations in VR and how this is applied in medical fields with a special focus on acupuncture.

\subsection{The Acupuncture Training and Treatment Process}
Acupuncture, a TCM practice dating back thousands of years, is founded on principles of the Meridian Theory~\citep{ulett1998traditional}, as outlined in the ancient text \textit{The Yellow Emperor's Classic of Internal Medicine}~\citep{veith2015yellow}. The two key concepts of this theory are meridian and acupuncture points (\aka acupoints). Meridian is a network of energy pathways that run throughout the body, where the life energy (called \emph{qi}) flows. Acupoints are the specific points lying on the meridian where the qi converges. Acupuncturists in ancient times healed patients through the proper stimulation of acupoints.

Adapted from traditional acupuncture, current Medical Acupuncture~\citep{white2009western} incorporates knowledge of anatomy, physiology, pathology, and the principles of EBM. 
%
Despite the unclear link between TCM acupoints-meridians and gross anatomy, the WHO standard \citep{white2004brief} uses modern anatomical terms to describe acupoints and meridians. Nowadays, medical acupuncturists often work closely with other healthcare providers to provide integrated care. This integration of Western and Eastern practices has allowed practitioners to develop new and effective treatment methods that target anatomical structures and nerve pathways directly to produce specific physiological effects, such as reducing pain or stimulating the immune system~\citep{white2009western}. As such, innovative acupuncture modalities arose, such as acupuncture for scalp~\citep{hao2012review}, eye~\citep{shao2020eye}, auricle~\citep{wang2001auricular}, and wrist~\citep{zhu2014wrist}.

Throughout history, acupuncture education has maintained a holistic approach while placing emphasis on cultivating a comprehensive understanding of the human body~\citep{veith2015yellow,white2004brief}. In modern times, acupuncture education institutes have further developed their programs to incorporate a more detailed understanding of human anatomy, physiology, and pathology, complementing the traditional holistic principles of ancient acupuncture practices~\citep{white2009western}. However, acupuncture education poses unique challenges, as it is still more conventional and traditional, and lacks some of the resources and technological advancements found in Western medicine~\citep{chen2019perspective}. Furthermore, the fundamental differences between TCM and EBM make it more difficult to integrate the diagnosis, treatment, and explanation of TCM and EBM in education~\citep{hong2018challenges}. 


A typical acupuncture treatment begins with an initial consultation and physical examination, during which the practitioner assesses the patient's health history and current symptoms. The practitioner then inserts thin needles into specific points on the patient's body to elicit the patient's sensation (called \emph{de qi}). The practitioner may further provide recommendations for self-care or follow-up appointments and adjust the treatment plan as necessary.
Using needles to stimulate specific anatomic sites is the core step in the treatment.
Strategic needle insertion into specific anatomic sites is a crucial component of acupuncture treatment. Although generally considered safe, even slight inaccuracies during needling can lead to unintended bleeding or organ puncture~\citep{vincent2001safety}. As a result, novice practitioners may struggle to locate and insert needles correctly, potentially resulting in subpar treatment or worse, patient safety. In addition, individual differences in anatomy and sensitivity to stimulation can make precise needling more challenging~\citep{lin2013exploration}.

\subsection{Virtual Reality in the Medical Field}
VR technology has been widely adopted in everyday's workplace~\cite{chen2023papertoplace, Chen2021ExGSense} and medical field for various applications, including diagnosis~\citep{liu2019review}, surgical planning~\citep{ayoub2019application,zhang2022directx,chheang2021collaborative}, and therapy~\citep{liu2022virtual,emmelkamp2021virtual}, offering new opportunities for training and education~\citep{nicholson2006can,pottle2019virtual}. 

With VR, medical professionals can interact with 3D models of CT scans, MRIs, and other medical images in a more immersive and intuitive way~\citep{pires2021use}. For example, visualization of 3D medical structures in radiotherapy treatment in VR significantly improves precision and reduces mental load and effort~\citep{chen2022exploring, chen2022vrcontour}. VR therapy has shown promising results in treating conditions such as \textbf{P}ost-\textbf{T}raumatic \textbf{S}tress \textbf{D}isorder (PTSD)~\citep{gonccalves2012efficacy}, chronic pain~\citep{jones2016impact,goudman2022virtual}, and offers progressive support with meditation~\citep{feinberg2022zenvr}. Advancements in VR have opened up new avenues for collaboration in many medical scenarios, allowing healthcare professionals to work together in immersive virtual environments~\citep{krauss2021current,vasilchenko2020collaborative}. For surgical planning, VR allows medical teams to understand the spatial relationships among internal structures and plan out complex procedures in advance~\citep{negrillo2020role,alsofy2020evaluation,chheang2021collaborative}. VR can also support emergency medical teams with interactive checklists and more broadly, dynamic cognitive aids~\citep{wu2013head}.

In education, VR has been shown to enhance performance and engagement, resulting in improved retention of information compared to traditional learning methods~\citep{krokos2019virtual, jensen2018review, radianti2020systematic}. Medical imaging in VR provides a detailed and accurate representation of anatomical structures, particularly in complex cases with nuance structures, such as cardiac anatomy in medical education~\citep{maresky2019virtual}. VR technology has been increasingly accepted to simulate patient-based procedures~\citep{maresky2019virtual, gunn2018use} and is recognized to improve cadaveric temporal bone dissection performance when compared with traditional teaching methods~\citep{zhao2011can}.

As acupuncture becomes increasingly important in the medical field, there is growing research and development focused on using VR technology to improve the effectiveness of acupuncture treatments in clinical settings~\citep{liang2021analysis}. Early-stage research constructed 3D digital human models integrating meridian and acupoints~\citep{heng2006virtual, kanehira2008development}. However, these systems and methods are primarily rooted in ancient theory and practice, which can limit the depth of anatomical understanding. To address these limitations, advanced VR training systems have been developed, leveraging detailed 3D anatomy for enhanced learning experiences~\citep{chen2019application}. For example, a study at the Guangzhou University of Traditional Chinese Medicine used a virtual acupuncture teaching system to demonstrate meridian outlines, acupoint anatomy, and needling techniques~\citep{rao2020practical}. 
One notable company that focuses on making VR solutions for general medical training experience is MAI~\citep{mai} which commercialized AcuMap~\citep{maiacumap}, the world's first VR training solution for acupuncture that incorporates anatomy, meridians, and acupoints on the virtual human model. Another recent trend in VR acupuncture research focuses more on exploring 3D user interactions made possible by the development of gesture recognition~\citep{zhang2017preliminary}, face alignment~\citep{zhang2021faceatlasar,zhang2022faceatlasar}, and full-body tracking~\citep{du2022mobile}, pushing the training scenarios from virtual reality to mixed reality~\citep{ryu2020design, yang2021development}. While recent acupuncture research has utilized x-rays, neuroimaging, and ultrasound to standardize the identification of acupoints~\citep{kim2015partially}, locate new needling sites~\citep{cao2020neuroimaging}, and investigate safety in terms of needling depth~\citep{lin2013exploration}, there is still a lack of widely available medical imaging visualization tools specifically designed for acupuncture treatment and research.

In summary, the existing research and commercial landscape shows how current VR products for acupuncture are primarily based on standard 3D anatomy models for training, which are adapted solely from EBM anatomy education, rather than co-designed with acupuncture domain experts. This creates an important gap and an opportunity to investigate whether these designs are suitable for acupuncture students and what affordances might be particularly beneficial in this setting. 
In addition, as modern medical acupuncture relies on evidence-based principles, it is important to also explore the principles that VR acupuncture applications (for education and clinic) should use to incorporate patient imaging data in addition to relying on standard 3D models.
Our need--finding study aims to understand the current needs and pain points in acupuncture training and practice, and explore the role that VR technologies can play in addressing existing challenges.  
Our overarching goal is to promote effective VR systems that integrate key human-centered design elements for an increasingly universal acupuncture practice that spans both TCM and EBM.
\section{Methods}

Our study is rooted in semi-structured interviews and participant observation. We first conducted observations of acupuncture treatment and tutoring workflows, and later engaged in semi-structured interviews with acupuncture students and doctors. This user-centered design~\citep{chamberlain2006towards} approach allowed us to facilitate an in-depth understanding of the barriers that acupuncture practitioners face in their daily routines. The study was approved by the Institutional Review Board~(IRB).

\subsection{Participant Recruitment}
\begin{figure*}
  \centering
  \includegraphics[width=\textwidth]{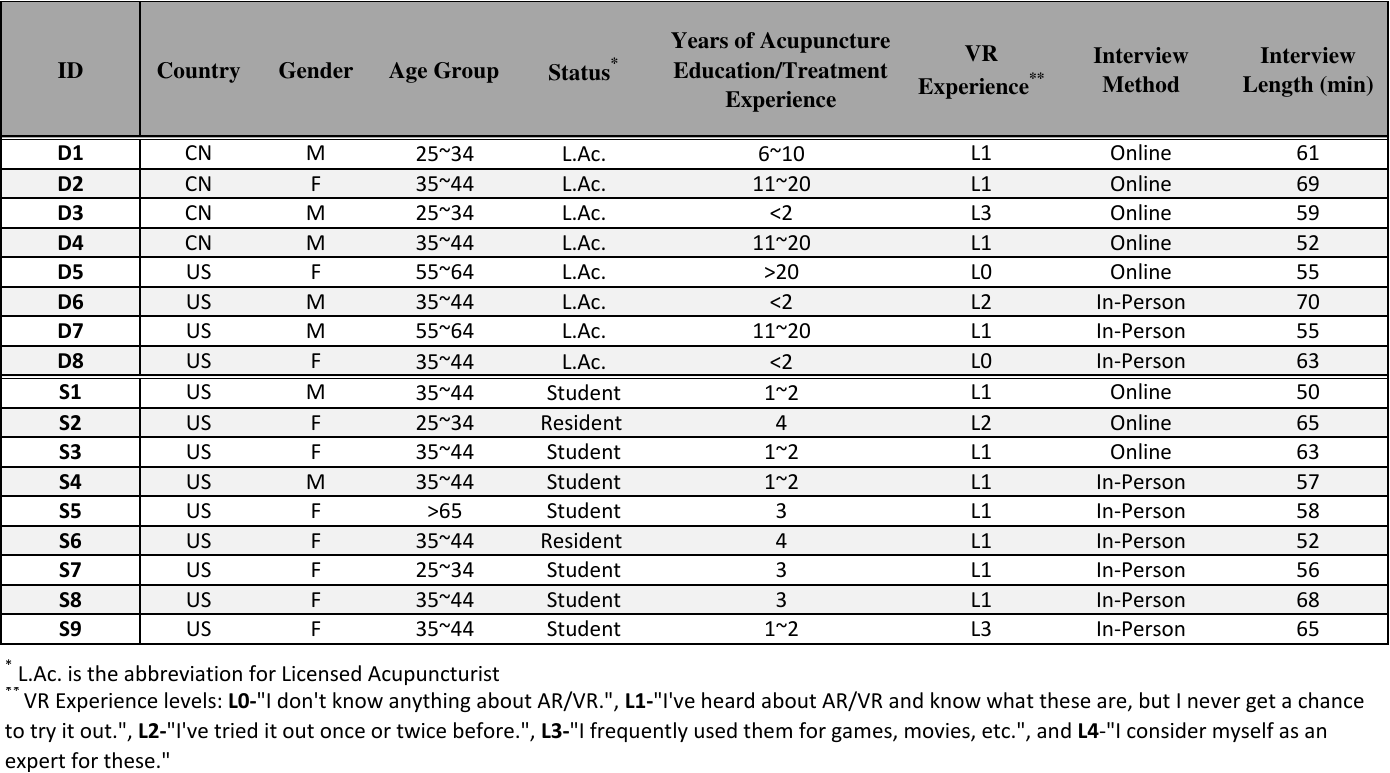}
  \caption{Demographics of the recruited participants, including eight acupuncture \emph{\underline{D}octors} and nine \emph{\underline{S}tudents}.}
  \label{fig:participants}
\end{figure*}

We recruited $17$ domain-expert participants with different experience levels, including eight doctors and nine students (see Fig.~\ref{fig:participants}). The participants were recruited from five professional acupuncture institutes, including three teaching hospitals, one private group clinic, and one individual clinic. 

The participants originated from China and the United States. China is where acupuncture first developed and has a rich history dating back thousands of years. Chinese acupuncture was the first form of TCM to actively seek integration into Western medicine~\citep{white2004brief}. In the United States, acupuncture is recognized as a primary care modality and is increasingly being incorporated into mainstream medical practices. Studying acupuncture across China and the United States provides a multi-perspective examination, and further allows us to investigate its effectiveness and potential applications in this context, resulting in a more comprehensive and holistic understanding of acupuncture treatment and education.

\subsection{Participant Observation}
The observations were conducted three months before the interviews in order to become familiar with the workflow of acupuncture, as well as to prepare questions for the interview round.
 We observed one treatment session in a private group institute in China and five treatment sessions in a teaching hospital in the United States, with both interns and licensed practitioners performing the acupuncture.
Two of the researchers also underwent acupuncture treatment to better understand the patient-side of the workflow. Most of the treatment sessions lasted about 50 minutes, with a few shorter ones lasting around 30 minutes. For a typical treatment observation, we entered the clinic room with the doctor's and patient's consent and stood by to observe their conversation and physical diagnosis. 
We were not allowed to record any images or videos, but we took notes of major events and interactions. When the doctor asked the patient to prepare for the treatment, we left the room with them to observe discussions with their colleagues and advisors. When time permitted, we also asked clarifying questions. Afterward, we followed the doctor back to the clinic room and observed them placing needles and administering treatments to the patient. Finally, when the session was over, we all exited the clinic room and discussed any lasting questions about the observation.

\subsection{Semi-structured Interviews}
We individually conducted interviews with all $17$ participants, with each interview taking $55$ - $65$ minutes. Since we wanted to also root our investigation into the potential of using VR immersive technology in acupuncture, we encouraged the candidates to participate in person so that we could expose them to a first-hand VR experience (as further explained in Sec.~\ref{sec:prob}).

\begin{figure*}
  \centering
  \includegraphics[width=\textwidth]{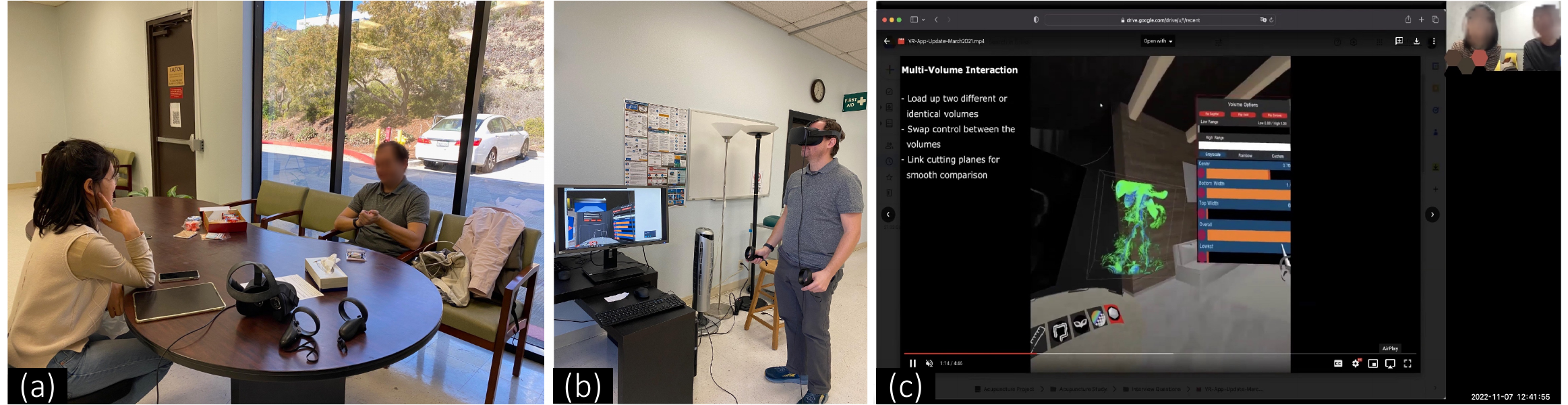}
  \caption{Settings of (a - b) in-person, and (c) online interviews. (a) Designer and participant in the conversation. (b) A participant is exploring a VR medical application as part of our cultural probe exposure. (c) During online interview, designers are illustrating the VR medical application over the pre-captured video as part of our cultural probe exposure.}
  \label{fig:interview}
\end{figure*}

\vspace{4px}
\noindent{\bf Pre-Screening Questionnaire ---}~
To address the complexity of acupuncture and its intersection with people's cultural, professional, and social backgrounds in the United States~\citep{chen2019perspective}, an online pre-screening questionnaire was developed. This enabled a more thorough understanding of the participants before conducting interviews, allowing for relevant topics to be discussed in greater depth. The questionnaire, detailed in Appendix~\ref{app:survey}, includes $12$ multiple-choice questions and aims at collecting the following data:

\begin{itemize}[leftmargin=*]
    \item Demographic information.
    \item Expertise level in acupuncture.
    \item Frequency of various collaborations in their daily routine.
    \item Use of medical imaging in their practice.
    \item Experience with VR/AR.
\end{itemize}

\begin{figure*}
  \centering
  \includegraphics[width=\textwidth]{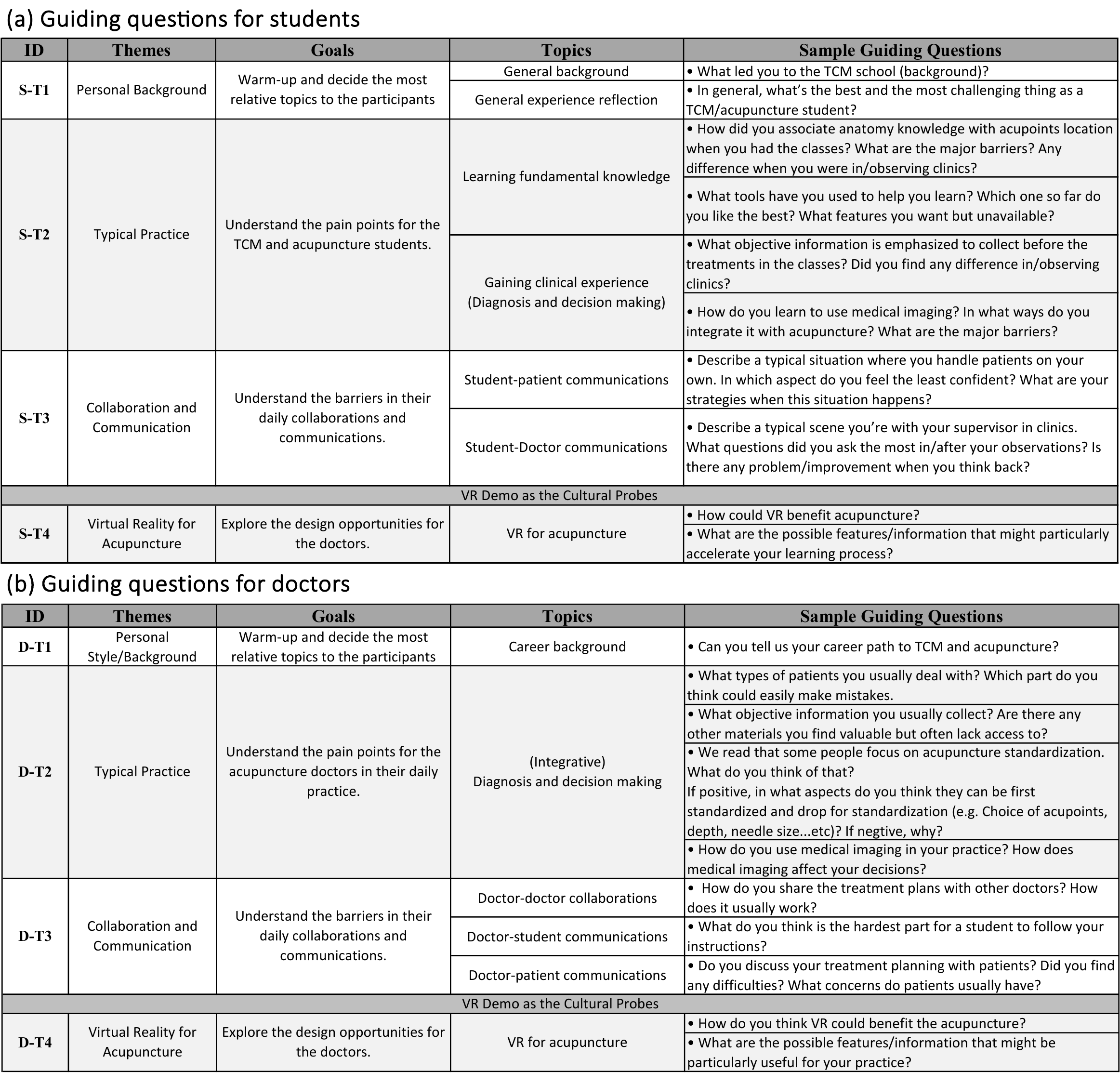}
  \caption{Guiding questions for students (a) and doctors (b).}
  \label{fig:interview_questions}
  \vspace{-1em}
\end{figure*}

\vspace{4px}\noindent{\bf Remote and In-person Interviews ---}
The topics for semi-structured interviews were divided into \emph{\underline{S}tudent} (hereafter noted as S) and \emph{\underline{D}octor} (hereafter noted as D) topics and were kept consistent within the group (see Fig.~\ref{fig:interview_questions}).  We collaboratively designed the interview topics with experienced TCM doctors, who helped us frame them in professional terms grounded in acupuncture practices. Since the participants had quite unique backgrounds, we encouraged them to tell us about their experiences by using open-ended guiding questions and adjusting our questions while adapting to the conversations.

All interview sessions were recorded with participants' permission and transcribed for further analysis. Then, two researchers analyzed all interviews using inductive coding approaches and generated specific codes from the collected data using grounded theory~\citep{strauss1994grounded}. We then examined and grouped the codes and concepts to describe the findings in Sec.~\ref{sec:finding}.

\vspace{4px}\noindent{\bf VR Cultural Probe ---}
\label{sec:prob}
In general, our study examined needs in acupuncture training and practices to later identify opportunities for new technologies. 
We also aimed to elicit some direct feedback from participants on VR as a possible new paradigm in acupuncture.
Given the participants' varying experiences with VR and AR (see Fig.~\ref{fig:participants}), we introduced a cultural probe~\citep{crabtree2003designing} to help familiarize participants with VR technology and inspire ideating around new design opportunities. To reduce bias in terms of need-finding, the cultural probe was only introduced at the end of the interviews.

\begin{figure*}
  \centering
  \includegraphics[width=\linewidth]{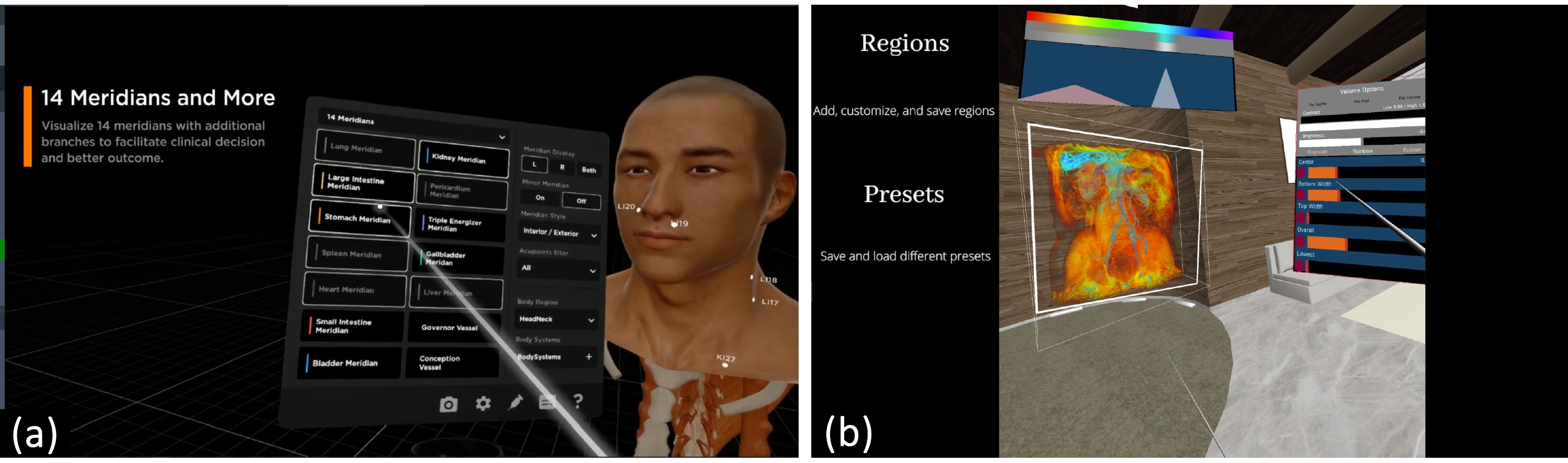}
  \caption{Cultural probe materials. (a) A screenshot of MAI's AcuMap demo video showing a partial standard 3D human anatomy model with selected meridians and acupoints. (b) A screenshot of the VR surgical planning application showing an anonymized MRI data volume with a visualization toolbox.}
  \label{fig:prob}
\end{figure*}

For our cultural probe, we showed MAI's AcuMap demo video~\citep{acumapvideo} and an existing VR application for surgical planning~\citep{zhang2021server} (see Fig.~\ref{fig:prob}). 
We first showed a footage of AcuMap's demo, usually 40 min into the interview. After an initial discussion over the video, we moved to the medical imaging-based surgical planning VR application~\citep{zhang2021server}.
For the in-person interviews, we first demoed the application while illustrating each functionality. We then allowed them to explore the visualization and 3D interaction in the application themselves. During the online interviews, we played a pre-recorded demo video and provided explanations for each operation and feature as it played.

Final discussions took place after the demo session. After the discussion, we asked the participants about their opinion and investigated their attitudes toward the VR experience for medical imaging. We later delved deeper into possible ways to utilize VR technology for acupuncture education and practice by asking specific questions related to VR for acupuncture (S-T4 or D-T4 questions).

\section{Results}
\label{sec:finding}

\begin{figure*}
  \centering
  \includegraphics[width=\linewidth]{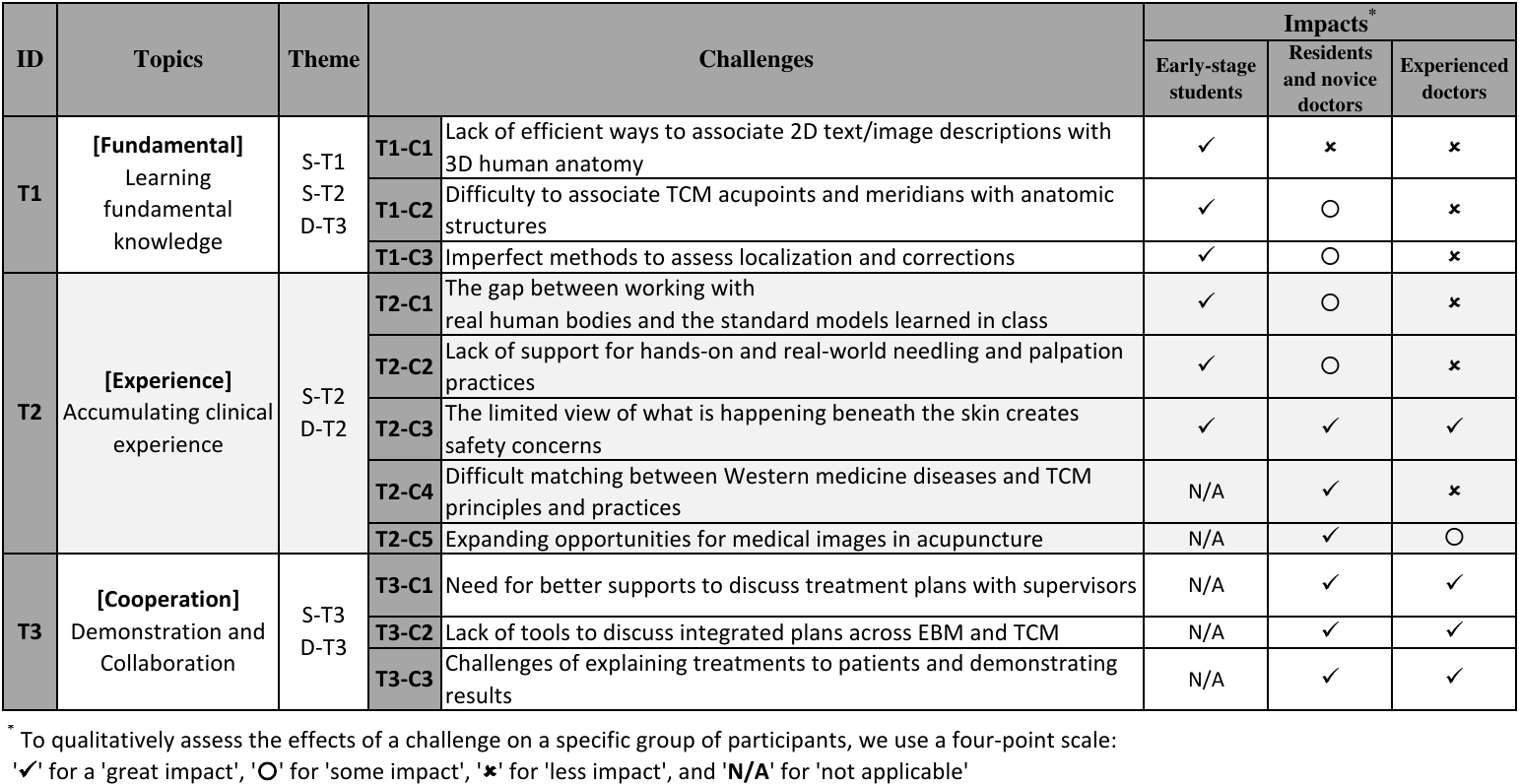}
  \caption{Overview of challenges emerging from acupuncture doctors and students across four categories. Refer to Fig.~\ref{subfig:ques-doc} and ~\ref{subfig:ques-stu} for the theme indices.}
  \label{fig:challenge}
\end{figure*}

In this section, we examine the challenges and needs that emerged during our need--finding study and the opportunities to introduce immersive VR technology to support acupuncture practice. Given the diverse profiles and needs of our participants, we will discuss findings from the perspective of (i) early-stage students, (ii) residents and novice doctors, and (iii) experienced doctors. Insights generated by specific doctors are referred to as [D1]-[D8], while we refer to students' insights using [S1]-[S9]. Fig.~\ref{fig:challenge} summarizes our findings and categorizes them into three themes: Learning Fundamental Knowledge (T1, Sec.~\ref{sec:T1}), Accumulating Clinical Experience (T2, Sec.~\ref{sec:T2}), and Demonstration and Collaboration (T3, Sec.~\ref{sec:T3}).

\subsection{Learning Fundamental Knowledge}
\label{sec:T1}

Almost all students described their first 1-2 years of experience as difficult and overwhelming, as 
%
they felt overwhelmed by the amount of information they had to memorize and understand, including knowledge of both TCM and EBM. 
%
%
%
We outline below four overall challenges in learning the fundamental knowledge.

\subsubsection*{\textbf{[T1-C1] Lack of efficient ways to associate 2D text/image descriptions with 3D human anatomy}}
During discussions, we observed that participants used their past learning experiences or re-enacted some of their clinical cases to explain their challenges. It was, however, interesting to witness how they struggled when trying to construct 3D anatomy structures in their heads to be able to explain. 
As an example, S4, a second-year student, highlighted that his current challenge is grasping the anatomical positioning of acupoints in relation to the first and second ribs. As a former nursing student, he felt that his previous education of the human body was too compartmentalized:
%
\begin{quote}
\emph{``I never considered everything so three-dimensional, like the depth of the skin and the tissue and layers, and thinking about how organs interact with one another.''}~[S4]
\end{quote}
Their supervisor, D6, an experienced acupuncturist, stressed the importance of {\it thinking in 3D} for early-stage students: {\it ``students need to think of the acupuncture points and the meridian system as a 3D concept, not 2D. How shallow the lung is, how deep the kidney is, and how many layers of muscles are there before the needling''}~[D6].

 
Despite the crucial role of understanding 3D concepts in acupuncture, many students still expressed reliance on 2D resources such as textbooks, including digital companion apps which were also limited to a 2D format. During the interview, S1 pulled a piece of paper with a stick figure of a foot, pinpointing several acupoints:{\it``I was practicing annotating the acupoints with my chart. It's like a coloring book, and that's what I usually do''}~[S1]. While some students reported benefits from using 3D anatomy education apps~\citep{anatomyatlasapp,3danatomyapp}, none were specifically designed for acupuncture learning: 
\begin{quote}
   \emph{``When it comes to TCM learning materials, they are just 2D illustrations and written descriptions.''}~[S6].
\end{quote}
Therefore, none of the participants were able to identify an application that effectively demonstrates acupoints and meridians in a 3D context.

\subsubsection*{\textbf{[T1-C2] Difficulty to associate TCM acupoints and meridians with anatomic structures}}


%
Accurately locating acupoints based on anatomy was identified as the biggest challenge among entry-level participants due to the separation of TCM and EBM, specifically the use of acupoints and channels in TCM, and the lack of a clear correlation between these concepts and modern anatomical understanding. For example, S8 expressed that {\it``the difference between how western medicine and TCM see organs [...] livers, kidneys, heart spleen are not necessarily organs referred to in western medicine"}~[S8]. S2 shared the same opinion and mentioned that {\it``learning them both is almost thinking with two minds, but they're never really separated"}~[S2]. Students with Western medicine backgrounds, Like S9, mentioned how {\it``[their] previous knowledge helps [them] ease some stress of the overwhelming components of this program''}~[S9]. However, as TCM emphasizes more on integrating the anatomical order and the organ system, linking new TCM knowledge with EBM concepts sometimes means {\it``rewiring [their] brain, because the mind was always thinking Western''}~[D5].

Furthermore, integration of EBM and TCM practice requires \textit{``a deep level of understanding beyond memorization, of the connection between the theories''}~[S8]. While such integration can be quite beneficial, today's educational support is not on par. Students perceive a significant disparity between the way knowledge is illustrated in EBM and the lack of support to learn TCM materials. As a student graduating soon, S2 could still vividly remember how she felt in her first year:
\begin{quote}
   \emph{``I remember oftentimes, I would close my eyes to visualize where the meridians would go on the body... So what I was missing was the visualization of the nerves and the meridians in a very clear way''}~[S2].
\end{quote}
To further investigate why it's so vital to see both the visible anatomical structures and invisible meridians, an experienced doctor, D1, who recognized the great importance of acupuncture standardization, explained: {\it``visualizing the relationship between acupoints and anatomical structures is crucial for achieving reproducibility in treatment plans''}~[D1]. S3, who grew up in a family with a long-standing reputation for practicing TCM, commented how it used to take a very long time for someone to become a skilled acupuncturist because of the lack of efficient ways to understand anatomy precisely, and further emphasized that {\it``the high precision correlation between acupoints and anatomy in acupuncture makes it efficient to cultivate remarkable acupuncturists''}~[S3]. 

\subsubsection*{\textbf{[T1-C3] Imperfect methods to assess localization and corrections}}

Early-stage students are trained with palpation skills to identify the acupoints on the surface skin before any needling practice, yet they often face challenges in determining the correct acupoint locations. They lack methods to assess the accuracy of their localization, which can impede their progress in training to become skilled acupuncturists. The main reason is the limited resources available in training. For example, S5 commented on challenges to be precise and improve acupuncture performance: {\it``it really takes a long time to get the point locations. You have to rely on the once or twice-a-week lab experience to know if you're on the right spots and have the memorization''}~[S5]. 

While gaining precision in acupoint localization requires practice, feedback is a crucial aspect of this process. Without regular engagement with experts and instructors, students lack the guidance and direction necessary to make continuous improvements in their skills. The limited interaction with experts and instructors can hinder the progress of students, as they are eager to learn faster and have access to interactive applications for support: {\it``I think it could be beneficial if it [3D Anatomy] could provide an interactive experience and people can play and learn''}~[S2], said S2, who is a big fan of using phone anatomy applications, but complained about their lack of interactive components. S4 confirmed how {\it``[they] use stickers and probes to practice, and how big a sticker is compared to a needle. Even so, it's still hard to tell if the stickers are on the right spots''}~[S4].

\subsection{Accumulating Clinical Experience}
\label{sec:T2}

\subsubsection*{\textbf{[T2-C1] The gap between working with 
real human bodies and the standard models learned in class}}
Exposure to the clinics as assistants in the early stage is believed to contribute heavily to students' performance. Students cherish their observation sessions as opportunities to engage in diagnosis and treatment procedures with experienced acupuncturists and begin to formulate their own assessments, as mentioned by S5: {\it``the observations let me know [that] the physical examination involves more than just identifying the anatomical location; it also involves understanding how the body feels and functions at various points''}~[S5]. While these sessions are helpful, some students outlined how they wished they could get additional hands--on practice:
\begin{quote}
   \emph{"There's such a long leading time to being able to actually work with real people in clinics, you don't [have the chance to] do it until much later. There's a limitation to how much you can do before then"}~[S4].
\end{quote}
While Western medical schools commonly introduce early--stage students to autopsy training as a way to improve their understanding of human anatomy and pathology, common TCM educational practices (both in the US and China) lack such resources and opportunities. As pointed out by D1, a doctor from China, \textit{``not every student will have a chance to attend a cadaveric training session''}~[D1]. Similarly, in the US, \textit{``only a small portion of students will consider taking a cadaver--based training program from a western medical institute after getting an acupuncture license"}~[D6]. 
One such example in our participants' cohort was a novice doctor, D8, who finished her cadaveric training program 1-2 years ago and further emphasized its importance, also pointing out how \textit{``from the experience of cadaver training, the most stunning fact is the variation [of circumstances and] of individuals.''}
In our interviews, we uncovered how it is common for students to be surprised by the difference between the real human body and the standard anatomical model they learned on, as confirmed by S8:
\begin{quote}
   \emph{``Everything we're learning in school is just basis and foundation. The clinic is very different from the textbook in class and your exams, because no patient is going to present themselves exactly like that.''}~[S8]
\end{quote}
Even current visual aids do not seem to help bridge the gap; for instance, when we showed participants the AcuMap video as a cultural probe, D8 pointed out how \textit{``it's as if there's space between these tissues. But it's not. It's vacuum sealed, and it's just entirely different. That's why the standard model cannot bring the whole picture.''}~[D8]

\subsubsection*{\textbf{[T2-C2] Lack of support for hands--on and real--world needling and palpation practices}}
Participants frequently discussed the concept of \textit{``experience"}, yet we noticed that the term held different meanings for different individuals. For experienced doctors, it mostly referred to treating patients with complex medical conditions,
%
while for novice doctors, gaining experience by seeing a large number of patients was deemed most crucial. Even with guidance from supervisors, proper localization of target points through palpation and accurate needle depth requires experience, as each patient's body and sensitivity to needling are unique.
Although students have lab classes to practice on each other as operation subjects, this opportunity is not sufficient: \textit{``The lab is great because you can palpate on your partner, but it's only still one body, one body structure''}~[S9]. To overcome this lack of practice, students usually ask their families and friends to act as additional operation subjects. S4 further confirmed the struggle of trying to learn from real bodies:
\begin{quote}
   \emph{``When I practice on my big-muscles neighbor, it's a totally different feel compared to my lab partner. Everybody's anatomy is unique, and it can be challenging to feel the right tenderness and decide the right depth''}~[S4].
\end{quote}
Such practice leads to insufficient feedback on whether the localization of the acupoints is correct, as previously discussed (see T1-C3 in Sect.~\ref{sec:T1}). However, the stakes are higher at this stage, as inaccurate localization may result in harm to the patient. 

\subsubsection*{\textbf{[T2-C3] The limited view of what is happening beneath the skin creates safety concerns}}

While acupuncture is generally considered a safe procedure, it might also cause life-threatening accidents~\citep{veith2015yellow}. For example, some acupoints, which are usually near important organs, nerves, or arteries, are referred to as dangerous or difficult acupoints. These points can be especially challenging and risky to needle if proper technique is not used.
Some participants reported preferring practicing on simpler points due to safety concerns, unless they're confident in the patient's anatomy. For example, D4 usually doesn't encourage his students to choose these difficult points, \textit{``unless [they] are an acupuncture master''}~[D4]. 
D3, a novice doctor also mentioned that he tends to play the safe card and stressed concerns about more difficult cases: {\it``safety always comes first, and then comes effectiveness''}~[D3].

Other participants also stated that proficient practitioners should be well--trained in school and \textit{``should be fearless of them [the difficult points]''}~[D5].
When we referred to the difficult points as \textit{``dangerous points''} while talking with D7, he chuckled and corrected us: \textit{``we could say they're `difficult points'. Still, they're not actually `dangerous'. Students are not very confident in clinics mostly because they don't highly understand the anatomy below the points''}~[D7]. Being trained in the same institute, S7 found it very important not to be afraid of any parts of the body. This is usually achieved by knowing the anatomy to a high degree. 

Aligned with prior work presented in literature review (Sect.~\ref{sec:background}), the participants pointed out the importance of imaging technology as an effective way to enhance acupuncture safety. 
After engaging with our cultural probe and reflecting on the power of visualization, almost all participants agreed that visualizing what happens beneath the skin would address the challenge of performing acupuncture on difficult points and improve effectiveness and safety. 
\begin{quote}
   \emph{``We take four years to practice safe needling techniques, yet we still avoid them during real practice... Sometimes it's necessary to do that [needling difficult points], and then it becomes a problem. [At this point] I want to take things out and see what's underneath''}~[S2].
\end{quote}
Similar to S2, students and novice doctors stated how better visualizations could help primarily to \textit{``understand how deep the needle is going''} and \textit{``to make sure you're not needling a nerve, or hitting something''}~[S9]. More experienced doctors also raised other reasons why visualization of the structure beneath the skin of specific patients could be a game changer. We summarize these reasons below across (1) balancing effectiveness and safety, (2) emphasizing patients' local anatomy, and (3) debriefing acupuncture sessions.

\begin{enumerate}[leftmargin=0.5cm]
\item \textbf{Balancing effectiveness and safety} --
As discussed above in T2-C2, practitioners with different levels of expertise can achieve different levels of effectiveness even with the same treatment plans. Supervisors noted that novices tend to needle shallower or find replacements on difficult points. D2 elaborated on this concept and how it relates to expertise and training:
%
\begin{quote}
   \emph{``Students usually don't feel confident about using a longer and thicker needle to stick deep enough for fear that it would cause injuries. But if the local lesion is deep, or the patient is not sensitive to needling, the treatment efficacy is insufficient in this way''}~[D2].
\end{quote}
\vspace{.5em}
  
She believed that students would feel secure and confident in targeting points when they can visualize the structures beneath the skin, which will lead to more effective acupuncture.

\vspace{.5em}
\item \textbf{Emphasizing patients' local anatomy} --
As introduced above in the background section (Sect.~\ref{sec:background}), medical acupuncture now involves techniques that directly needle the local anatomy of the patient, which brings both opportunities and challenges for needling. 
%
As an example, D6 mentioned that there is a technique in musculoskeletal treatment in which the needles target a specific joint or area of musculoskeletal discomfort: 
\textit{``that's when the patient's medical imaging data should be vital''}~[D6].
Regardless of how a target point is identified, the disease and individual characteristics of each patient can add additional complexities to the local anatomy in acupuncture treatment. D7 made a specific example:
\begin{quote}
   \emph{
``Asthma patients use their muscles for breathing, which changes the anatomy structures and makes the `difficult points' even more difficult. 
   So it's very important to learn the accurate situation [for the particular patient]''}~[D7].
\end{quote}
%
%

\vspace{.5em}
\item \textbf{Debriefing acupuncture sessions} --
Additional visualization has the potential to be a unique method to further reflect on the acupuncture practice after the treatment sessions. We noticed how the idea of introducing any extra device during a treatment session was usually perceived negatively by the practitioners, given how it might add burdens for both the doctors and patients. However, most of our participants mentioned the potential of also seeing patients' bodies right after the treatment, as argued by D8:
\begin{quote}
   \emph{``[After the treatment,] I would be able to relate what I'm sensing internally with a visual image. I would be able to inform and learn from those images to improve my needling technique because I would have feedback''}~[D8].
\end{quote}
\end{enumerate}

\subsubsection*{\textbf{[T2-C4] Difficulty connecting Western medicine diseases and TCM principles and practices}}
\label{sec:t2c4}

While TCM and Western medicine are distinct, they are often used in conjunction in China and the US. TCM doctors may collaborate with Western medical doctors to provide comprehensive care. Therefore, it is important for TCM doctors to match their practices with Western diagnoses and treatment plans for effective care. 

Though this seems straightforward, in practice, it can be challenging especially for students and novice TCM doctors who 
lack experience with Western diagnostic criteria. Our findings in this context can be summarized around three main pain points: (1) TCM diagnoses lack objective standards, (2) it is difficult to distill effective objective and subjective information, and (3) it is challenging to make integrative treatment plans.

\vspace{-.5em}
\begin{enumerate}[leftmargin=0.5cm, listparindent=1em]
\item \textbf{TCM diagnoses lack objective standards} --
Unlike EBM diagnoses, TCM diagnoses rely more on subjective judgments than objective measures, meaning that they rely on the practitioner's observations and interpretations rather than on measurable data. 

For example, pulse diagnosis~\citep{walsh2007pulse} is one of the four major assessments, which involves using fingertip pressure to feel a patient's pulse through the skin. Many participants admitted that even experienced doctors may struggle with pulse diagnosis.
S3 commented on the potential risks of relying on the subjective diagnoses of TCM: \textit{``it's hard to convince the patients to trust TCM. For certain patients, it's crucial to explain your decision-making process and show the treatment plans with evidence''}.

\vspace{.5em}
\item \textbf{Difficulty distilling and communicating patient's objective and subjective information} --
During the participant observation stage, we observed three treatments provided by residents under the supervision of their supervisors. 
Generally, the practitioners first gather patients' symptoms, conditions, and medical history.
Alongside, they examine tongues, take the pulse, and perform a physical examination to formalize treatment plans. Each diagnostic technique confirms and validates the diagnosis. Then, they brief their supervisors on the patient's condition to ensure that the treatment plan is optimized, while the patients prepare and rest. This is usually a very concise communication, as \textit{``[they] have to build [the skills of] how to present a case to a supervisor properly...however, [the residents] may not get it well"}~[S2]. 
Supervisors expressed similar concerns about students: \textit{``[students] only ask patients' symptoms as a process but not actually extracting the crucial information for diagnosis''}~[D7].

Besides, acupuncture treatment sessions often involve longer communication and more diagnostic information than common primary care visits. This can be both beneficial for learning about patients and challenging for highlighting crucial information. S8 recounts one particular episode:
\begin{quote}
   \emph{"I'm listening to my patients and want to share all the same [with my supervisor]. But there are so many... My strategy now is to stick to their chief complaint and the supporting facts. What actions can relieve the symptoms immediately"}~[S8].
\end{quote}

\vspace{.5em}
\item \textbf{The challenges to create integrative treatment plans} --
Our interviews outlined how the integration of TCM and EBM methods can enhance learning.
%
%
For example, as we discussed the difficulties mastering subjective TCM pulse diagnosis, S2 explained her strategy on \textit{``[using] Western medical diagnosis to aid in understanding different pulse presentation."}


More generally, experienced doctors agreed that for common diseases, incorporating Western medicine with Eastern diagnosis provides a valuable learning opportunity for students. They identified one major challenge in the integration of these two approaches, with Western medicine focusing on local anatomical stimulation (referred to as \textit{branch}) while TCM emphasizes whole-body, systemic treatment based on TCM theory (referred to as \textit{root}). 
The participants underlined that the most effective diagnosis and treatment plans integrate both root and branch methods in order to achieve optimal results. However, under such complex cases, D1 mentioned how \textit{``diseases that are visible to the naked eye are actually quite easy to diagnose. But those that are interrelated and functional are more difficult. For example, a patient with lower back pain is likely not just a problem with the lumbar spine [...] it could be an issue with the digestive system, or it could be a problem with the kidneys"}.
\end{enumerate}

\subsubsection*{\textbf{[T2-C5] Expanding opportunities for medical images in acupuncture}}
\label{sec:t2c5}

Given the importance of patient anatomy in acupuncture treatments, doctors often recognize medical imaging as a critical dimension of data. In clinics, patients, especially those with orthopedic issues, usually bring their medical images to their acupuncture doctors. With these data and reports from the radiologists, \textit{``their findings will help us tailor the treatments to be more specific"}~[D7].
D8, for instance, recalled how her special training in medical imaging contributes to her acupuncture practice now: 
\begin{quote}
   \emph{``The way imaging contributes [...] it really helps me see the bone and connective tissue, the facet joints, nerves, and what's going on to the disk space."}~[D8]
\end{quote}
For this reason, she emphasized the value of access to medical images of the same disease in different individuals during the training stage, stating that this is the \textit{``golden thing''} [D8] they desire most. Similarly, D4 outlined that medical images are valuable resources for teaching students about pathology and objectively explaining treatment plans. He believes that the fastest way for a student to learn is through imitation and practical application. To this end, D4 envisions using a set of images from different patients with the same disease to demonstrate the difference in both EBM and TCM theories, and how he can tailor treatment accordingly.

Although acupuncture doctors recognize the value of patient medical images, our pre-screening questionnaire revealed that they are not being utilized as frequently as needed. We found that this is a common issue both in China and in the US, often due to a lack of equipment or insurance coverage. As a result, doctors in both countries tend to only refer patients for medical imaging when palpation diagnosis indicates a serious concern that may impact the decision for acupuncture treatment. D1 comments on this issue and outlines how, if it were possible, adding medical imaging to the current workflow would be ideal:
\begin{quote}
   \emph{``My workplace does not have the necessary conditions for taking medical images. So if the patient has not had an imaging examination before, I usually will not require them to go to the hospital for an examination before treatment, unless it is particularly severe. I can diagnose from a physical examination. But for objectivity and also some legal requirements, it is still necessary to do so. If we have the equipment nearby, we will certainly do it. I am inclined towards this more clear and objective diagnosis. This is also much easier to be understood (by the patients)"}~[D1].
\end{quote}
\subsection{Demonstration and Collaboration}
\label{sec:T3}

Demonstration and collaboration activities include explaining treatments and demonstrating the treatment results. In this section, we outline opportunities that arise from such collaborative activities in three common scenarios:
\begin{itemize}
    \item A student working with their supervisors as observers or assistants in clinics.
    \item A doctor explaining the treatment to the patient and their families.
    \item A TCM doctor collaborating with Western doctors.
\end{itemize}
The themes below outline demonstration and collaboration activities, cutting across all three scenarios.

\subsubsection*{\textbf{[T3-C1] Need for better supports to discuss treatment plans with supervisors}}

In the previous section, we outlined how presenting observations and treatment plans is challenging for residents (T2-C4). During those exchanges, we sometimes observed confusion and a breakdown of the communication, as was also confirmed by S6:
\begin{quote}
   \emph{``I'm not always on the same page with my supervisor. When this situation happens [a disagreement], I usually follow their recommendations and think about it later'' }~[S6].
\end{quote}
Reflecting on similar episodes, their supervisors also recalled that the most difficult part is helping students understand the reasoning behind choosing one acupoint over another, as students have difficulties understanding the basis of each selected point. S4 believes that using interactive visualization in VR could solve this problem:
\begin{quote}
   \emph{``I can see it integrated so well if we had a tool like that [Virtual Reality]. Imagine we have a clinic group, and then in the centers where we meet with our supervisor, we have the visual to where we come out and talk, where you could show them the 3D imaging of the body. It can be used for [practical] technique-wise purposes, but also, maybe show what happens when I push over here and how it affects the other side of the body or something''}~[S4].
\end{quote}
Besides, student S6 wanted the application to further show different types and sizes of needles to help her decide which needle to use and how far different sizes of needles would have to go into the spine and muscles: \textit{``our supervisors don't usually talk about such information much''}~[S6].

\subsubsection*{\textbf{[T3-C2] Lack of tools to discuss integrated plans across EBM and TCM}}
As we mentioned above, more and more doctors realize the importance of collaborating across Western and Eastern approaches. 

For example, as expressed by D6, it is beneficial for both TCM and EBM doctors to work together: \textit{``We're not at a point where it's easy to request labs and imaging. Knowing how to network and talk to other professional practitioners, MDs, and physical therapists is a good practice. It helps build relationships, network, [submit] cross-referrals, and consult in difficult cases. It would be a win-win''} [D6]. 

A good example of such collaboration is the rehabilitation clinic where D2 works: a significant part of her work in post-stroke rehabilitation is collaborative, as she described when talking about the rehabilitation assessment meeting they regularly hold for the incoming patients: 
\begin{quote}
   \emph{``Our assessment meeting includes rehabilitation therapists, rehabilitation physicians, rehabilitation nurses, traditional Chinese medicine practitioners, massage therapists, and physical therapists. The patient's attending physician will lead the discussion, and each person will report from their own perspective, after which we will have a comprehensive discussion. As a TCM doctor, I need to explain the diagnostic principles, the acupoint selection criteria and reasons, herbs, and additional treatments such as massage to everyone. Helping the patient achieve the functional level they want to recover requires the work of the entire team and a holistic approach.''}~[D2].
\end{quote}
While this meeting is highly collaborative, D2 believes that there is a need for better support to discuss patients' data and understand the treatment plan. A similar example raised by D6 is post-surgery recovery, in which acupuncture has been demonstrated as an effective rehabilitation method: \textit{``We need to better understand where and what is happening with the surgery, where the scar tissue might be, what channels are affected''}~ [D6].

Overall our participants stressed that
an effective treatment plan should be based on understanding each other's practice and \textit{``Being able to work integratively will help patients recover better''}~[D7]. But currently, there is a lack of suitable tools to facilitate seamless collaboration and communication.

\subsubsection*{\textbf{[T3-C3] Challenges of explaining treatments to patients and demonstrating results}}

New providers in the medical field often find it challenging to feel confident when interacting with their patients in clinical settings. 

D3 mentioned that a significant challenge for him is the need for confidence: as a new doctor who has just begun working independently without the guidance of supervisors, he lacks confidence in his abilities. Additionally, even when dealing with confident cases, he sometimes has hesitation about communicating with patients. 
To dive deeper into this barrier, when participants discussed their lack of confidence, we asked about their strategies for handling this situation. Many mentioned that being able to provide an intuitive illustration of their practice could help reduce uncertainty and build trust with patients. For example, S2 believes that her ability to intuitively present a TCM treatment plan in conjunction with the patient's EBM diagnosis instills trust. Also, from D2's experience, effectively demonstrating the progress made in the treatment so far is the most effective way to build reliance with patients and their families: \textit{``it is best to provide data or image-based stimuli to the family members. For example, looking at an image and saying, Ah, one-third has healed, and two-thirds on the second time! This kind of intuitive effect is easily recognized [...] if you use professional methods to explain the process and emphasize the effect, most people can accept it''}~[D2]. 

D1 explained how a major barrier to introducing such visual support is that TCM treatment relies on patients' or practitioners' subjective feelings, which are \textit{``not always reliable for building the whole picture''}~[D1]. He further raised an example about some studies using tomography to visualize the effectiveness on the needling region, \textit{``if I had such a device in clinics, I'd also use it for demonstration''}~[D1].


%
\vspace{1em}
Based on the feedback and challenges reported by acupuncture doctors and students, it is clear that incorporating immersive visualization tools into their daily routines could have a significant impact. In the next section, we will explore the specific functionalities that should be designed to meet their needs.

\section{Discussion}
In this section, we first summarize how the challenges raised by the participants are rooted in different acupuncture tasks. We then discuss how specific functionalities can be implemented to support participants' requirements for each task. 

\subsection{Major Tasks in Acupuncture Requiring Additional Support}
\begin{figure*}
  \centering
  \includegraphics[width=0.8\linewidth]{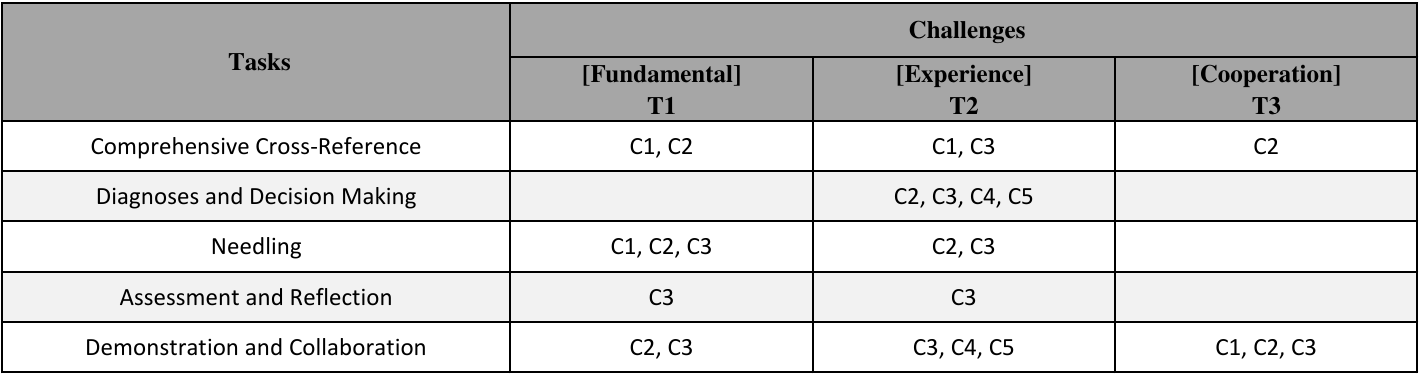}
  \caption{The five major acupuncture-related tasks that need additional support, and the list of challenges that these need to overcome. The full description of each challenge can be seen in Fig.~\ref{fig:challenge}}
  \label{fig:task-challenge}
\end{figure*}

Through our need--finding, we identified five major tasks, for both doctors and students, that need to be supported to overcome the challenges that they have reported (Sect.~\ref{sec:finding}). To better outline the connection between the tasks, the themes, and the challenges outlined above, in the discussion below, we refer back to specific Themes with (T\#) and Challenges with (T\#-C\#). Fig.~\ref{fig:task-challenge} summarizes the five major tasks we identified and the challenges that they need to overcome from our findings.

\vspace{+.05in}\noindent
{\bf (1) Comprehensive Cross-Referencing} --
The main task of the first 1-2 years of study is to learn the fundamental knowledge (T1) of both TCM and EBM. A Cross-Reference that provides a comprehensive overview of these two approaches to medicine, and helps students understand how to integrate them into their practice would be highly beneficial. As a tool to build theoretical foundations, it can cover a wide range of topics, including TCM acupoints and meridians (T1-C1), and the ability to associate TCM and EBM body theories (T1-C2). A Comprehensive Cross-Reference can also serve as a valuable reference for students to consult as they continue to learn and develop their skills as practitioners. If the cross-reference is augmented with individual body references, it can also provide guidance on how to apply this knowledge in clinical settings (T2-C1, T2-C3, and T3-C2).

\vspace{+.05in}\noindent
{\bf (2) Diagnosis and Decision Making} --
Accumulating experience in diagnoses and decision-making is crucial for the professional development of TCM doctors. Despite having a solid theoretical foundation, residents and novice doctors may still encounter challenges in practice. This is because patients may present different TCM constitutions for the same western disease (T2-C4), and a lack of experience in palpation and needling (T2-C2) can make it difficult to determine the details of treatment plans such as needle size, angle, and depth (T2-C3). Although we expect doctors to keep building their skills and knowledge throughout their careers, one effective way of supporting the development of diagnostic and decision-making skills is to expose them to a wide range of medical images, both during training and during clinical practice (T2-C5).


\vspace{+.05in}\noindent
{\bf (3) Needling} --
Due to safety considerations, entry--level students only have limited opportunities to needle human bodies. To properly prepare for needling patients soon, they must develop the ability to locate target points precisely and utilize the correct needling techniques (T1-C1, -C2, and -C3). This requires a professional level of understanding of local anatomy structures and simulation of the needling practice. After properly building their abilities on standard anatomy models, the next challenge will be adapting to different individuals while maintaining a balance between effectiveness and safety. This is especially important for complex cases where a patient's previous therapy has affected local anatomy structures (T2-C2). In these situations, a full awareness of local anatomy structures is crucial to ensure effective and safe treatments (T2-C3).

\vspace{+.05in}\noindent
{\bf (4) Assessment and Reflection} --
Immediate assessment and feedback on the practice are crucial yet difficult to achieve in acupuncture training due to the subjective nature of the assessment process. While hands-on training and feedback can help students learn to apply their theoretical knowledge in a practical setting and receive immediate feedback on their progress, objectively assessing the correctness of their techniques is difficult. Often, students can only practice their techniques with the guidance and supervision of their instructors, which may not fully reflect the real-world situations they will encounter as practitioners (T1-C3). To overcome these challenges, acupuncture students need access to various assessment tools and methods that can accurately and efficiently evaluate their skills and knowledge. By actively seeking out opportunities for assessment and feedback, acupuncture students can continually improve their skills and confidence as practitioners (T2-C3).

\vspace{+.05in}\noindent
{\bf (5) Demonstration and Collaboration} --
Effective collaboration between doctors and students requires clear and efficient communication.
To support such collaboration tasks, especially for entry-level students, the clinical practice should include ways of helping novices to fully understand the TCM and EBM principles and techniques (T1-C2) and assessing their performance (T1-C3).
%
It is essential to guide students on making acupuncture plans for individual patients based on their medical materials because this allows them to accurately identify and target specific areas of concern. Medical imaging, such as MRI or CT scans, can provide detailed information about a patient's anatomy and physiological processes, which can be invaluable in the planning and execution of acupuncture treatment. Lacking such resources slows the students' progress in analyzing and interpreting medical data (T2-C5), matching an EBM disease with a TCM diagnosis (T2-C4), and their abilities to tailor the treatments for individuals (T2-C3).
Due to the differences between TCM and EBM training and overall philosophy, TCM doctors tend to be weaker in interpreting EBM materials, while EBM doctors have problems understanding TCM principles. TCM doctors often rely on a holistic approach to treatment, while EBM doctors tend to focus on interventions based on scientific evidence, and this can lead to differing perspectives on patient care. 
Clinicians lack an effective method to create a common ground between the two practices from both theoretical and technical perspectives (T3-C1, -C2, and -C3).

\subsection{Design Opportunities of VR-based Acupuncture}
\begin{figure*}
  \centering
  \includegraphics[width=\linewidth]{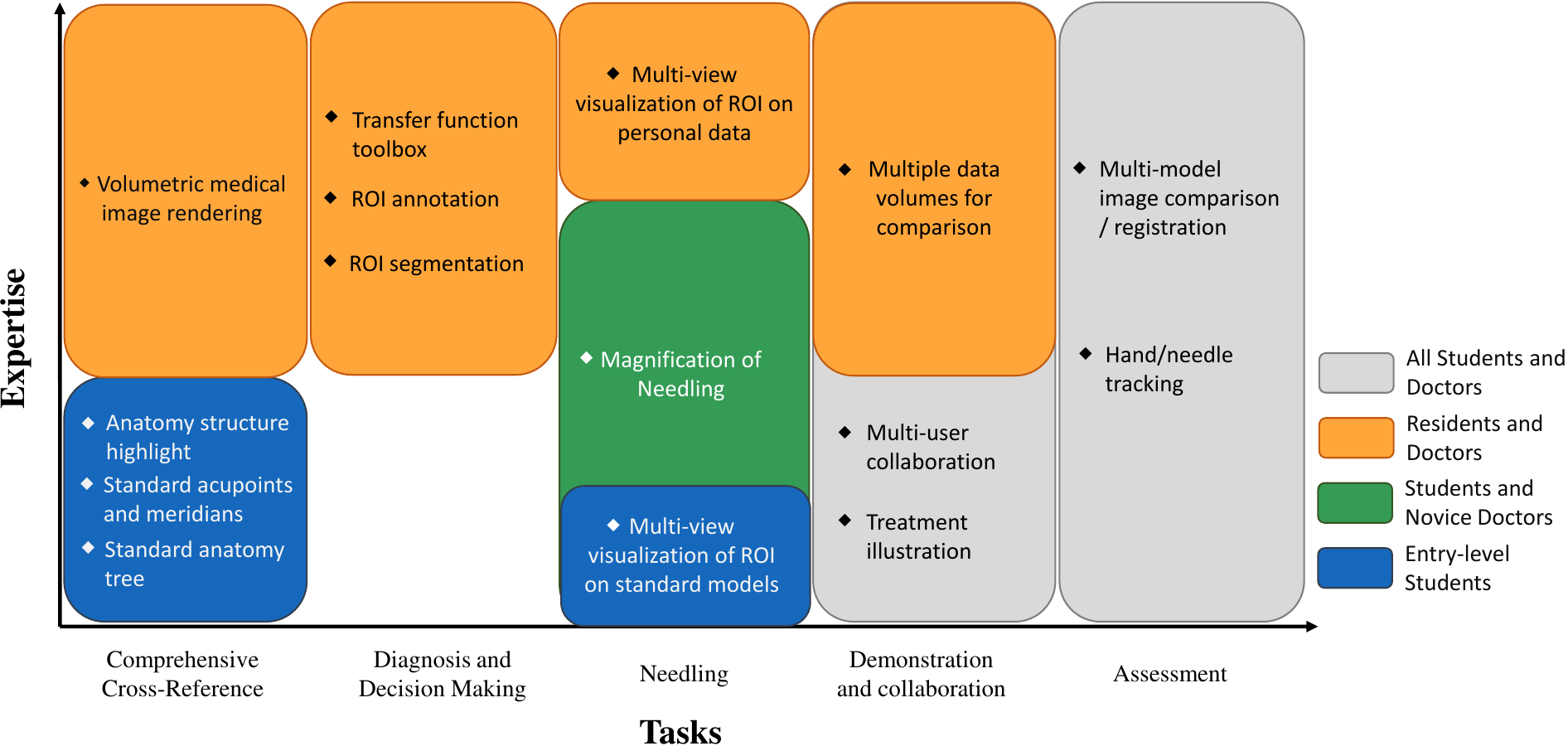}
  \caption{Technology features in five tasks for different levels of expertise acupuncture students and doctors to address their challenges.}
  \vspace{-1em}
  \label{fig:solution}
\end{figure*}

Based on our analysis of the tasks and challenges, we have identified two potential strategies that can cater to the needs of all three levels of expertise among our target users (Early-stage Students, Residents and Novice Doctors, and Experienced Doctors). Fig.~\ref{fig:solution} depicts the features of these two design strategies, outlining how they support the above-defined tasks (and hence overcome the challenges), and what level of expertise they support. 
Since our investigation was rooted in opportunities for immersive visualizations in Virtual Reality, we show how VR is the optimal strategy to implement both design strategies and their specific features. The two strategies can be implemented in a single VR application or across two VR applications, and users can choose the one that best suits their current needs.


\vspace{1em}
\noindent
\textbf{Design Considerations for the Integration of Standard Human Anatomy Features}\\
\noindent
As introduced in the related work, existing VR training systems for acupuncture visualize standard 3D human anatomy models marked with acupoints and meridians and already demonstrated effectiveness in existing research. In this section, we outline how tools that are used in these settings should be designed and extended to support the task outlined above and overcome the challenges reported by our participants.

Given the importance of learning as precisely as possible the location of acupoints and the meridians from TCM, and their correspondence with human anatomy from EBM, systems and visualizations should be able to support early-stage students guided by their supervisors. We identified the following features as core functionalities in this design: 

\begin{itemize}[leftmargin=*]
    \item \textbf{Visualization}
        \begin{itemize}
            \item Found on a detailed atlas of the anatomy of every single organ system.
            \item Overlay meridians and acupoints against human anatomical systems of the body.
            \item Highlight a meridian and show the flows.
            \item Provide an information panel to illustrate detailed information for each acupoint.
        \end{itemize}
    \item \textbf{Needling Simulation}
            \begin{itemize}
            \item Simulate needles of different types and sizes.
            \item Show an adjustable magnified view of the intersecting point with selected local anatomical components.
            \item Provide optional needling targets and assess the correctness of the current practice.
            \item Illustrate potential effects around the needling region.
        \end{itemize}
    \item \textbf{Collaboration}
        \begin{itemize}
            \item Allow multiple users to view and interact in the same virtual environment.
            \item Enable different levels of permissions for supervisors and trainees.
        \end{itemize}
\end{itemize}

\begin{figure*}
  \centering
  \includegraphics[width=\linewidth]{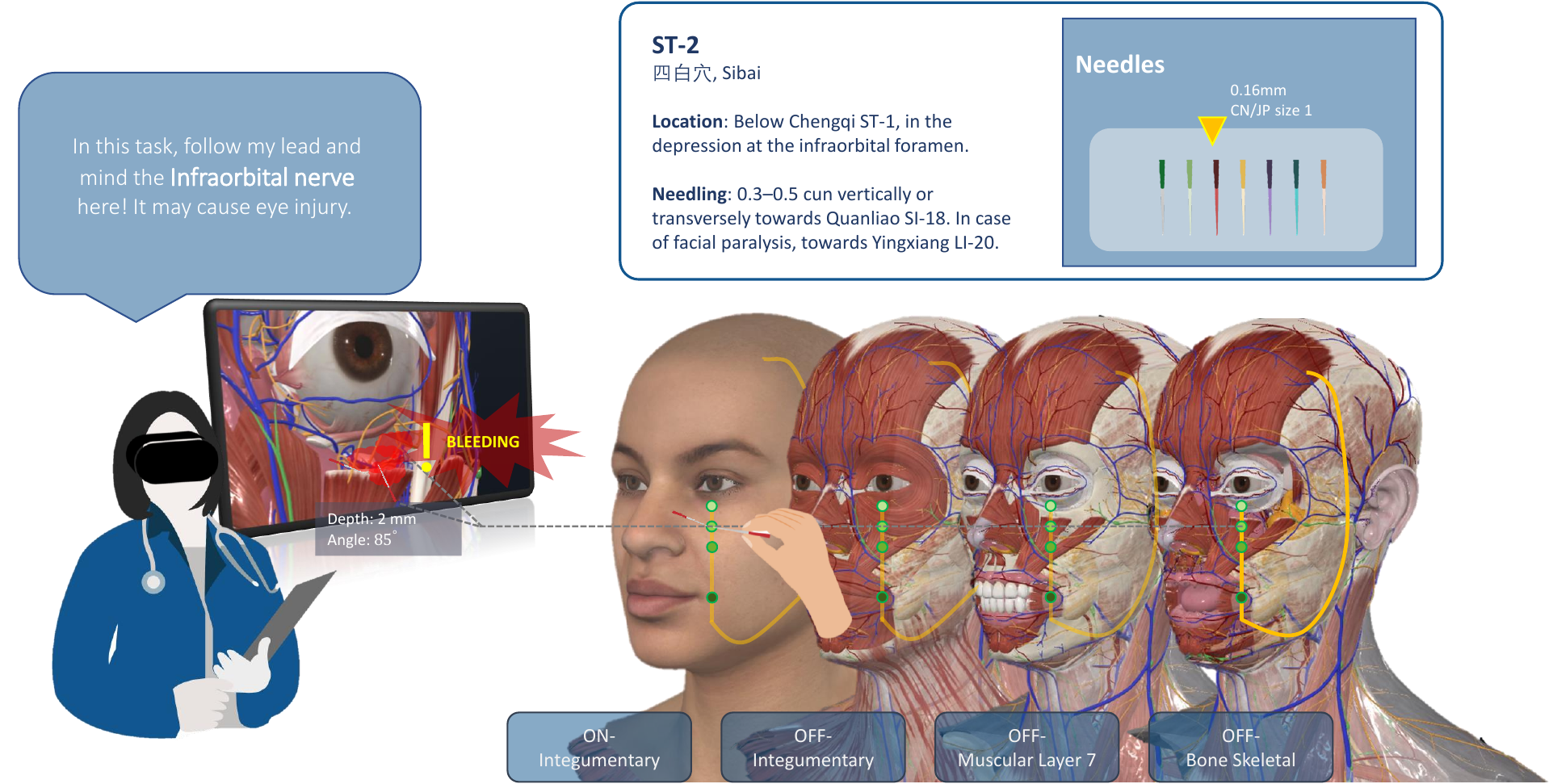}
  \caption{Design illustration for the Integration of Standard Human Anatomy. A trainee's view of needling acupoint ST2 on a standard human anatomy model guided by a supervisor inside the virtual training environment (Standard anatomy model images are captured from Elsevier's Complete Anatomy 2023~\citep{anatomyelseviersapp}, Windows version).}
  \label{fig:design-standard}
\end{figure*}

Combining these core functionalities will allow for the creation of immersive 3D visualizations that bridge the gap between 2D textbook information and 3D human anatomy through 3D rendering (T1-C1). Overlaying acupoints and meridians will directly outline the relationship with the local anatomical structures instead of showing information floating over the surface skin (T1-C2). Fig.~\ref{fig:design-standard}, illustrates a simple prototype showing how these features could be integrated in the trainee's view when working with a supervisor inside a virtual training environment. 

As a training application, besides providing a basic magnified view from the needling position, the core features above will provide ways for trainees to mimic and evaluate their choice of needles, needling angles, depth, and manipulations (T1-C3). For example, the acupoint ST2, which is situated close to the optic nerve and requires very high localization precision, is difficult to assess from the outside skin (T2-C3). As shown in Fig.~\ref{fig:design-standard}, an effective assessment tool would help with the localization, but also show the consequences of mistakes, like if the needle goes through a capillary, the body starts to bleed.
Finally, in order to facilitate collaboration, an efficient training system will not only support multi-user functionality but also implement a flexible permission system that accommodates the varying roles and responsibilities of users. For instance, as demonstrated in Fig.~\ref{fig:design-standard}, a supervisor should be able to assign and monitor tasks, while a trainee should be able to access and complete them. This will ensure that the users in different roles can work together effectively (T3-C1).

\vspace{1em}
\noindent
\textbf{Design Considerations for the Integration of Medical Imaging}\\
\noindent
As discussed above, training systems that only focus on standard human anatomy models are most effective in addressing the challenges of learning fundamental knowledge (T1). However, while standard models may be useful in certain contexts, they are inadequate for senior students to acquire clinical experience (T2) or for doctors to use in their clinical practice and communication (T3). To the best of our knowledge, few acupuncture training systems are based on visualization and manipulation of medical images.

Medical imaging techniques can include CT and MRI scans for preoperative procedures, and ultrasound or fMRI for intraoperative purposes. These techniques provide valuable information and insights that can help novice acupuncture doctors to gain clinical experience and to improve their diagnostic and treatment skills. Compared to a standard-model-based of training, based on our need-finding, a medical imaging-based planning and training system should provide the following functionality:

\begin{itemize}[leftmargin=*]
    \item \textbf{Visualization}
        \begin{itemize}
            \item Show high-resolution volumetric visualization from specific 3D medical images.
            \item Integrate a library of existing medical images from a public database.
            \item Visualize target needling points against human anatomical systems of the body.
            \item Provide the ability to observe the target regions of interest (ROI) with cross-section and 3D selection views.
            \item Introduce multiple data enhancement tools to pull out ROI.
            \item Enable automatic anatomical structure (e.g., organ, muscle, and nerve) segmentation and multi-modality image registration.
        \end{itemize}
    \item \textbf{Needling Simulation}
            \begin{itemize}
            \item Include all the needling simulation features as described for the standard-model-based one.
        \end{itemize}
    \item \textbf{Collaboration}
        \begin{itemize}
            \item Include all the needling simulation features as described for the standard-model-based one.
            \item Provide an annotation toolbox to type notes and make measurements.
        \end{itemize}
\end{itemize}

\begin{figure*}
  \centering
  \includegraphics[width=\linewidth]{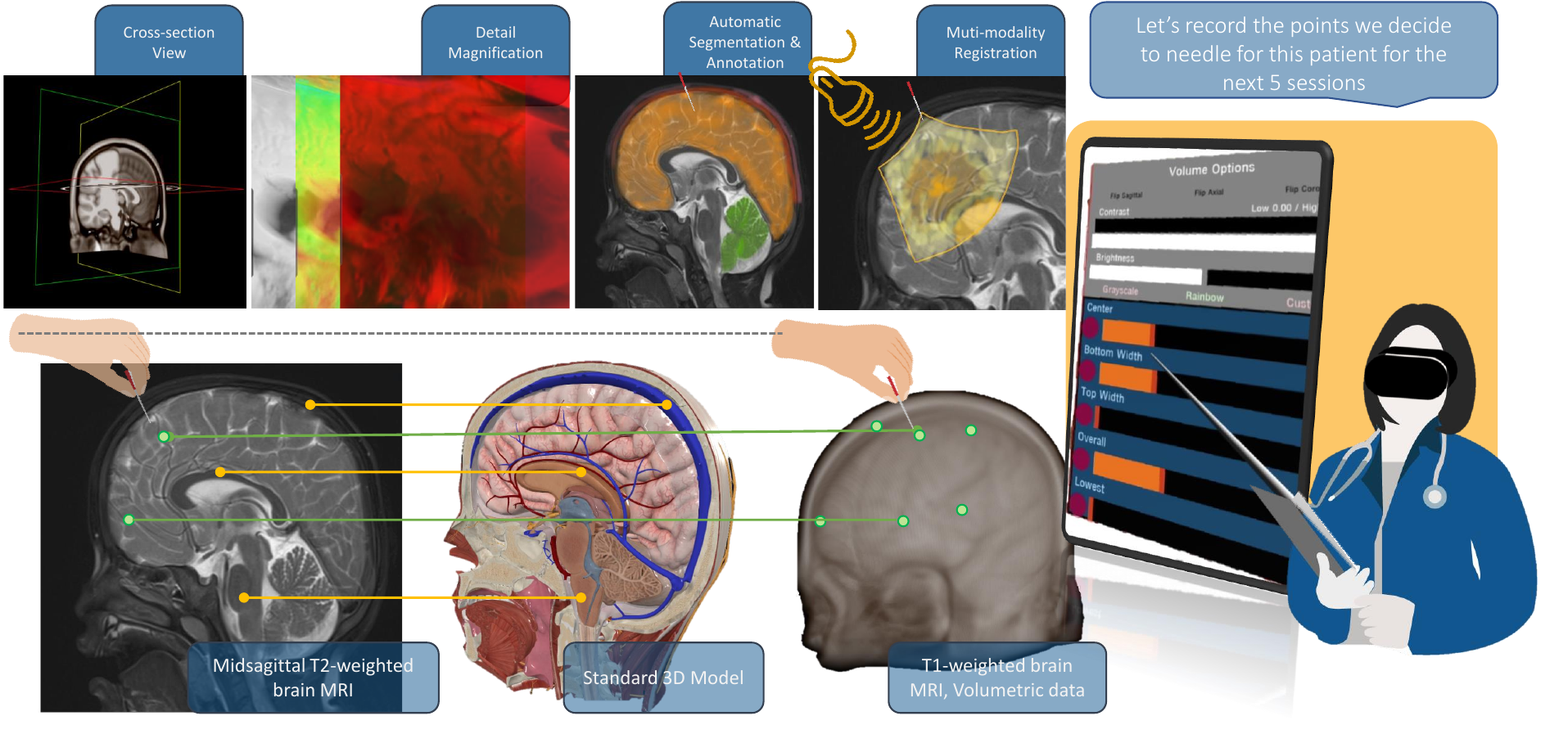}
    \vspace{-2.5em}
  \caption{Design illustration for the integration of medical imaging into acupuncture training. Shown is a resident's view of creating treatment plans with a supervisor inside the virtual environment, using a patient's MRI and Ultrasound scans (Sample brain MRI scans are from~\citep{ds004401:1.0.1} and Elsevier's Complete Anatomy 2023~\citep{anatomyelseviersapp}. Volumetric data visualization is from~\citep{socrvolume})}.
  \label{fig:design-medical-imaging}
  \vspace{-1.5em}
\end{figure*}

In Fig.~\ref{fig:design-medical-imaging}, we illustrate a prototype that implements such features in VR. Specifically, we envision a resident able to create treatment plans with a supervisor within a virtual training environment. The process utilizes the patient's brain MRI and ultrasound scans as the primary source of information, with the use of a standard model as a reference point for discussion.
This system serves as a dual-purpose tool for both acupuncture training and treatment planning. 
%
The core functionalities would allow students and doctors to enhance their training by bridging the gap between entry-level memorization of standard models and actual human bodies (T2-C1). Actual patients' images from patients and from a public database would give clinicians the opportunity to accumulate more experience in interpreting different diseases and then associate it with the TCM diagnosis and acupuncture treatments (T2-C4). The availability of viewing the same disease of multiple individuals will also help to address how different categories of TCM symptoms and constitutions are linked to the same biomedical situation (T2-C5). 

In addition, providing a comparison of healthy and unhealthy scans for each type of disease will help to visually demonstrate the impact of the disease on the body and potentially allow also patients to see what the clinician is thinking and alleviate a lot of their fears (T3-C1). 
During actual treatments in the clinic, when patients bring their own digital images, the clearer and more effective viewing of patients' body will enhance the objectivity of treatment decisions (T2-C4). Also, a complete understanding of the patient's situation inside their body will allow doctors to avoid potential scar tissue during the procedures (T2-C3). Additional functionalities like anatomical structure segmentation can also be introduced to help addressing the limited number of TCM doctors proficient in interpreting medical images, to manage the significant demand for assistance with identifying and segmenting ROIs, and to help localize the needling and explain some of the patients' pains through additional information for muscle and nerve segments.

Overall, an intuitive visualization system based on medical images would enhance communication between students-supervisors, patient-doctors, and TCM-EBM doctors. While EBM doctors widely accept medical imaging-based visualization systems for diagnosis, monitoring, and surgical planning, it could become a great starting point for TCM and EBM doctors to work together and bridge the gap between western and eastern medicine (T2-C4, T3-C1, and T3-C2). Repeated imaging pre- and post-treatment will allow a better discussion of the effects of acupuncture and will be essential to adjust the plans for the next step (T3-C3).

\subsection{Limitations}
In our study, we recognize that the level of experience with VR technology can impact participants' perceptions of its usefulness and potential applications. We acknowledge that only a small percentage of our participants had significant experience with VR, which may have limited the depth of their feedback on the technology. In future studies, we plan to address this limitation by providing more opportunities for participants to become accustomed to working in VR to ensure that they can provide more informed feedback on the technology.
\section{Conclusion and Future Work}

This paper presents a need--finding analysis that investigated the potential of immersive visualizations in VR to empower medical acupuncture doctors and students. The participants (n=17 acupuncture doctors and students) shared their experiences with learning, practicing, educating, and collaborating in their daily routines and discussed their challenges and design opportunities in these activities. We outlined those challenges at three different levels of expertise and identified three guiding themes: learning fundamental knowledge, accumulating clinical experience, and demonstration and collaboration. 

As the need for integrating the treatment rooted in traditional practice with modern technologies increases, it is important to identify the starting point to leverage medical imaging in acupuncture treatment. By reporting 11 barriers, we unveiled key opportunities for designing effective VR systems, and describe them in terms of their relationship to five specific tasks: comprehensive cross-reference, diagnosis and decision-making, needling, assessment and reflection, and effective collaboration, including interaction across TCM and EBM. We showed how two specific design strategies rooted in the visualization of standard human anatomy and medical imaging can empower acupuncture practitioners in their daily routine through state-of-art VR visualization and interaction technologies. 
We believe that the findings presented in this work can guide future researchers who aim to develop 3D visualization-based applications in the field of acupuncture and integrating medical processes into novel VR systems.

%

\bibliographystyle{ACM-Reference-Format}
\bibliography{references}

\appendix
\section{Pre-Screening Questionnaire}\label{app:survey} 
\begin{figure}
  \centering
  \includegraphics[width=\linewidth]{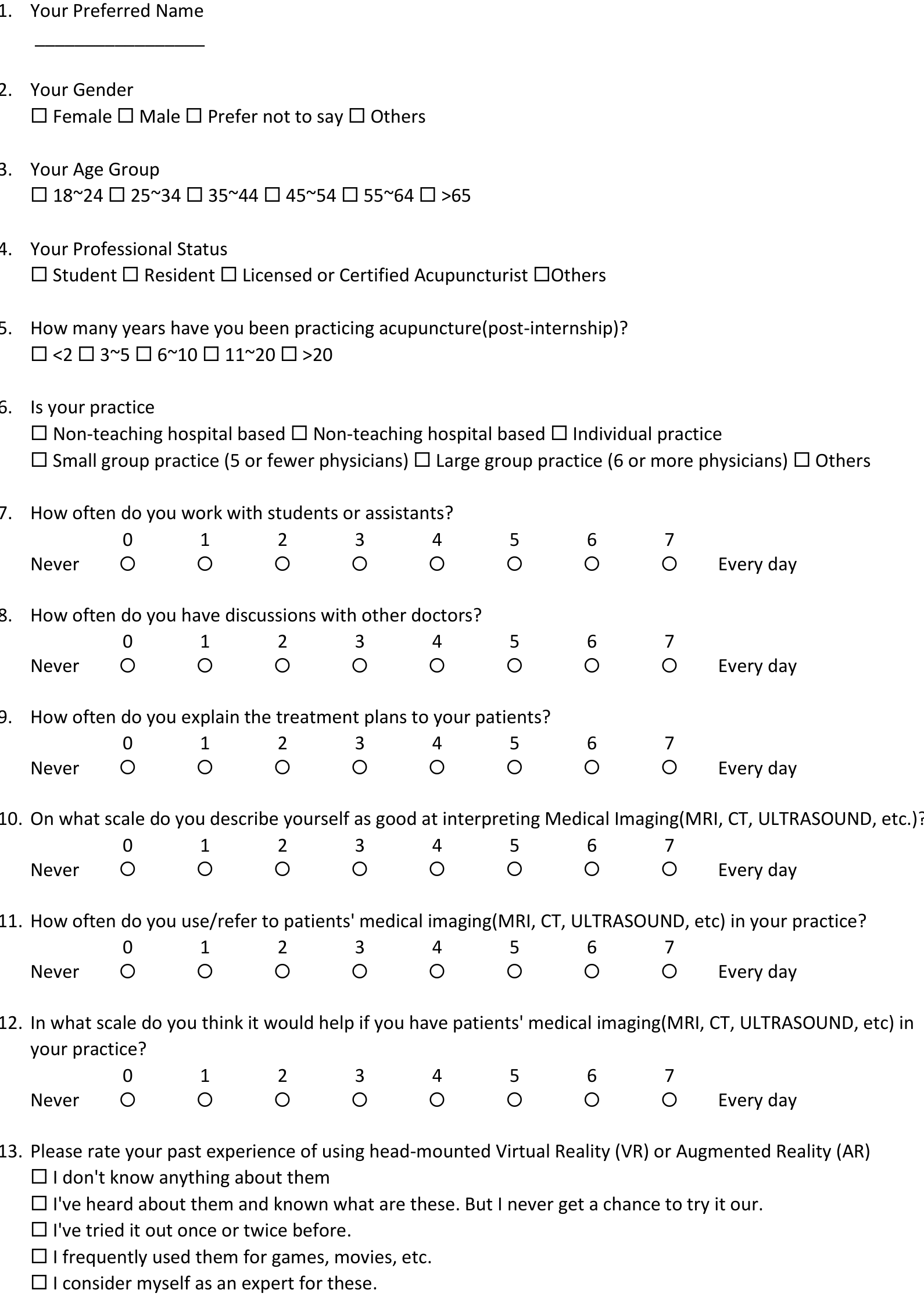}
  \caption{The 13-question pre-screening questionnaire.}
  \label{fig:pre-screening}
\end{figure}

This section provides the full pre-screening questionnaire for the participants (see Fig.~\ref{fig:pre-screening}).

\end{document}